\documentclass[prd,tightenlines,nofootinbib,prd]{revtex4}
\usepackage{amsmath,amscd,amssymb}
\usepackage{color,epsfig}
\usepackage{latexsym}
\usepackage{mathrsfs}
\usepackage{graphicx}
\usepackage{bbm}      
\usepackage[normalem]{ulem}

\def\beq{\begin{equation}}
\def\eeq{\end{equation}}
\def\be{\begin{eqnarray}}
\def\ee{\end{eqnarray}}
\newcommand{\dslash}{\partial \hskip -0.6em /}
\newcommand{\Dslash}{D \hskip -0.6em /}

\newcommand{\tr}{\mbox{tr}}

\newcommand{\Vek}[1]{\mbox{\boldmath$#1$\unboldmath}}

\newcommand{\zr}[1]{\mbox{\hspace*{#1em}}}
\newcommand{\ID}{\mbox{{\sf 1}\zr{-0.16}\rule{0.04em}{1.55ex}\zr{0.1}}}

\begin{document}

\title{Isospin Invariance and the Vacuum Polarization
Energy of Cosmic Strings}

\author{H. Weigel$^{a)}$, M. Quandt$^{b)}$, N. Graham$^{c)}$}

\affiliation{
$^{a)}$Physics Department, Stellenbosch University,
Matieland 7602, South Africa\\
$^{b)}$Institute for Theoretical Physics, T\"ubingen University
D--72076 T\"ubingen, Germany\\
$^{c)}$Department of Physics, Middlebury College
Middlebury, VT 05753, USA}

\begin{abstract}
We corroborate the previously applied spectral
approach to compute the vacuum polarization energy of string configurations in models 
similar to the standard model of particle physics. The central observation underlying
this corroboration is the existence of a particular global isospin 
transformation of the string configuration. Under this transformation the single 
particle energies of the quantum fluctuations are invariant, while the
inevitable implementation of regularization and renormalization requires 
operations that are not invariant. We verify numerically that all such variances 
eventually cancel, and that the vacuum polarization energy obtained in the 
spectral approach is indeed gauge invariant.
\end{abstract}

\maketitle

\section{Introduction}
Various field theories suggest the existence of stringlike configurations, 
which are the particle physics analogs of vortices or magnetic flux tubes in 
condensed matter physics. These configurations can arise at scales ranging from 
the fundamental distances in string theory to astrophysical distances, where in 
the latter case they are often called {\it cosmic  strings}. (See for example 
the reviews~\cite{Copeland:2011dx,Hindmarsh:2011qj}.)\footnote{Arguments for 
a closer connection between cosmic and fundamental 
strings are given in Ref. \cite{Copeland:2009ga}.} A well--known 
representative is the Nielsen--Olesen vortex~\cite{Nielsen:1973cs} in a model 
with an Abelian gauge field coupled to a single Higgs field. This vortex 
is classically stable, as are particular embeddings in non--Abelian models 
with several Higgs fields~\cite{Hindmarsh:2016lhy}. In general, however,
non--Abelian string configurations are not classically stable. In this 
context the $Z$--string, which typically involves the $Z$--boson field 
in the standard model, is of particular interest~\cite{Achucarro:1999it}.
Though not classically stable, it is possible that
these strings are stabilized by quantum effects. The vacuum polarization
energy (\emph{VPE}), which is the regularized and renormalized sum of
all zero point energies of the quantum fluctuations in the classical
background, is central to these investigations. In field theory
quantum effects are typically estimated by Feynman diagram
techniques. However, stringlike configurations have a nontrivial
structure at  spatial infinity which makes the formulation of a
Feynman perturbation expansion impossible without any further
adaptation. Even then, the convergence of the series is not guaranteed
as the relevant couplings are not  necessarily small and the series is
only asymptotic.  On top of that, the rich topological
structures~\cite{Kibble:2015twa} of theories with cosmic strings
require techniques beyond perturbative treatments. Not surprisingly,
the study of the \emph{VPE} of cosmic string configurations has a long
history  of slow progress without a fully concluding answer.

Numerous publications have analyzed quantum fluctuations about cosmic
strings. Naculich~\cite{Naculich:1995cb} has discussed that in the
limit of weak coupling, fermion fluctuations tend to destabilize the string.
The quantum properties of $Z$--strings have also been connected to
nonperturbative anomalies~\cite{Klinkhamer:2003hz}. Furthermore, the
emergence or absence of exact neutrino zero modes in a $Z$--string background 
and the possible consequences for the string topology were investigated 
in Ref.~\cite{Stojkovic}.  A first attempt at a full calculation of the 
fermion quantum corrections to the $Z$--string energy was carried out 
in Ref.~\cite{Groves:1999ks}. Those authors were only able to compare the 
energies of two string configurations, rather than comparing a single 
string to the vacuum, because of limitations arising from the nontrivial 
behavior at spatial infinity.  (We will discuss this issue in more 
detail below.) The fermion vacuum polarization 
energy of the Abelian Nielsen--Olesen vortex~\cite{Nielsen:1973cs} has 
been estimated in Ref.~\cite{Bordag:2003at} with regularization limited 
to the subtraction of the divergences in the heat--kernel expansion. 
On the other hand, quantum energies of bosonic fluctuations in 
string backgrounds were calculated in Ref.~\cite{Baacke:2008sq}.  
However, these are suppressed compared to fermion fluctuations 
when the number of internal degrees of freedom, {\it e.g.} color, is 
large.

Using the spectral method~\cite{Graham:2009zz} the (one--loop) \emph{VPE} can be 
computed from scattering data. An essential feature of this method is the 
identification of elements from the Born expansion with Feynman diagrams. 
These elements are added and subtracted to make contact with standard 
renormalization techniques and conditions which prescribe certain Green 
functions for particular values of transferred momenta. In a sequence of projects 
we succeeded computing the fermion \emph{VPE} of cosmic strings after solving a number 
of problems:\footnote{See Ref.~\cite{Weigel:2015lva} for a recent review.}
\begin{enumerate}
\item The string configuration does not have a well defined Born series to be 
identified with the Feynman series of quantum field theory;
this can be overcome by a special local gauge transformation~\cite{Weigel:2010pf}. 
\item A correction factor to the na{\"\i}ve Jost function is required to maintain 
the analytic properties of scattering data~\cite{Weigel:2009wi,Graham:2011fw}, 
because the effective fermion mass depends on the distance from the string core. 
\item Higher order Feynman diagrams are required which become exceedingly 
difficult to evaluate numerically; this is solved by the 
so-called fake boson approach~\cite{Farhi:2001kh}.
\end{enumerate}

\noindent
Formally the unregularized and unrenormalized fermion \emph{VPE} of the string 
is the sum of energy eigenvalues from a Dirac Hamiltonian. These eigenvalues 
are invariant for a particular path in the space of parameters which define the
(weak) isospin orientation of the string~\cite{Klinkhamer:1997hw}. Previous 
calculations of the \emph{VPE}~\cite{Weigel:2010pf,Weigel:2009wi,Graham:2011fw} were 
restricted to a simplifying submanifold in isospace that could not access this path. 
The invariance of the single particle energies is, however, not sufficient to ensure 
that the full fermion \emph{VPE} is also invariant in this calculation. The sum of the energy 
eigenvalues is ultraviolet divergent and in the inevitable process of 
regularization and renormalization divergent contributions emerge that are 
manifestly variant. They are conjectured to cancel based on their formal equivalence 
as expansions in powers of the string background. On the regularization side 
terms from the Born expansion to scattering data are subtracted, which on the 
renormalization side are added back in the form of Feynman diagrams. 
An exact match of these quantities is not at all obvious. 
For instance, Feynman diagrams allow us to distinguish between the divergences that 
emerge from the quantum loops and the Fourier modes of the background 
(this is e.g.~essential for understanding the Casimir effect~\cite{Graham:2003ib} in the 
context of spectral methods). On the other hand scattering data, and thus the Born expansion 
terms, do not distinguish between external and loop momenta.
Using dimensional regularization, the equivalence of the two schemes has been 
verified for the leading (tadpole) divergence, both for boson~\cite{Farhi:2000ws} and
fermion\cite{Farhi:2000gz} fluctuations. At higher order the distinction between
loop and Fourier momenta is essential and so far no such proof has been provided.

The scattering data decouple into angular momentum channels. As we will explain in 
Sec.~\ref{sec:vpe}, a channel by channel subtraction is mandatory for contributions 
that can be related to the \emph{quadratic} ultraviolet divergences in the Feynman series. 
The subleading logarithmic divergences require to include higher order Born/Feynman terms, 
which are very cumbersome to simulate numerically. Fortunately, the set of divergences 
terminates at this logarithmic level so that these divergences can be cavalierly treated 
by simulating them in a simpler (typically bosonic) theory. This method brings into the 
game an additional contribution that is not manifestly invariant under the particular 
isospin transformation mentioned above.  Furthermore, the simulation of divergences by 
a boson model also requires the exchange of momentum integrals with orbital angular 
momentum sums.  which by itself demands care: for instance, swapping these operations 
for momenta on the real axis gives erroneous results~\cite{Pasipoularides:2000gg}; instead, 
an analytic continuation to imaginary momenta is required~\cite{Schroder:2007xk}. 
In any event, the whole regularization procedure is not manifestly gauge invariant while
gauge invariance should, of course, be maintained by the final result in order for the 
adopted calculational procedure to produce unambiguous results. A good example to 
demonstrate the subtleties of gauge invariance in the context of the spectral approach 
are the \emph{vacuum charges} induced by a nontrivial background configuration: 
improper regularization may falsely predict anomalous vacuum charges~\cite{Farhi:2000gz}. 
From these considerations, it is clear that consistency checks are indispensable to ensure 
that the spectral method does not artificially break (gauge) symmetries leading to 
erroneous results. In the present paper, we will explore such a test based on a global 
isospin invariance. Because of the operation under item 1) above, this also probes
a local invariance.

\bigskip\noindent
We conclude this introduction with a brief description of our model.
The bosonic part is described by the Lagrangian
\begin{equation}
\mathcal{L}_{\phi,W}=-\frac{1}{2} \tr
\left(G^{\mu\nu}G_{\mu\nu}\right) +
\frac{1}{2} \tr \left(D^{\mu}\Phi \right)^{\dag} D_{\mu}\Phi
- \frac{\lambda}{2} \tr \left(\Phi^{\dag} \Phi - v^2 \right)^2 \,,
\label{Lbosonic}
\end{equation}
where the Higgs doublet is written using the matrix representation
\begin{equation}
\Phi=\begin{pmatrix}
\phi_0^* & \phi_+ \cr -\phi_+^* & \phi_0 \end{pmatrix} \,.
\label{higgsmatrix}
\end{equation}
The gauge coupling constant $g$ enters through both the covariant derivative
$D_\mu = \partial_\mu - i \,g W_\mu$ and the $SU(2)$ field  strength tensor 
\begin{equation}
G_{\mu\nu} = \partial_\mu\,W_\nu - \partial_\nu W_\mu - i \,g\,[ \,W_\mu\,,\,
W_\nu\,]\,.
\label{fieldtensor}
\end{equation}
The classical potential has been chosen such that the Higgs field 
acquires a vacuum expectation value (\emph{VEV}) $v$, where 
$\langle {\rm det}(\Phi)\rangle = v^2 \neq 0$. As a consequence, 
all bosons become massive: $m_W = g v / \sqrt{2}$ and $m_H = 2 v \,\sqrt{\lambda}$.
The interaction of the (classical) string with the left--handed fermions is described
by
\begin{equation}
\mathcal{L}_\Psi=i\overline{\Psi}
\left(P_L \Dslash  + P_R \dslash \right) \Psi
-f\,\overline{\Psi}\left(\Phi P_R+\Phi^\dagger P_L\right)\Psi\,.
\label{gaugelag}
\end{equation}
Here, $P_{R,L}=\frac{1}{2}\left(1\pm\gamma_5\right)$ are projection
operators on left/right--handed components, respectively, and the strength
of the Higgs-fermion interaction is parametrized by the Yukawa coupling $f$,
which gives rise to the fermion mass, $m = f v$.

This short report is organized as follows. In Sec.~II we discuss the particular
form of the cosmic string configuration and describe the 
path in weak isospace along which the Dirac eigenvalues are unchanged. In section III 
we explain how spectral methods are utilized to compute the fermion contribution to the 
\emph{VPE}, including the subtleties needed to make the approach feasible. 
We present numerical results for the \emph{VPE} in section IV and show that 
this particular invariance is indeed reproduced within our numerical accuracy. 
We conclude with a brief summary in Sec.~V and leave some technical 
details to appendixes.

\section{Cosmic String Configuration}

The starting point to parametrize cosmic string configurations is the four 
dimensional unit vector~\cite{Klinkhamer:1994uy,Graham:2006qt}
\begin{equation}
\hat{\Vek{n}}(\xi_1,\xi_2,\varphi)=\begin{pmatrix}
{\rm sin}\xi_1\,{\rm sin}\xi_2\, {\rm cos}\varphi\cr
{\rm cos}\xi_1 \cr {\rm sin}\xi_1 \,{\rm cos}\xi_2\cr
{\rm sin}\xi_1 \,{\rm sin}\xi_2\, {\rm sin}\varphi
\end{pmatrix}\,,
\label{eq:nhat}
\end{equation}
where $\xi_1$ and $\xi_2$ describe the isospin orientation of the string
and $\varphi$ is the azimuthal angle in coordinate space.\footnote{The string 
configuration will be infinitely extended along the $3$-direction in 
coordinate space.} For simplicity, we will always consider unit winding of 
the string; generalizations to winding number $n$ merely require the replacement
${\rm cos}\varphi\to{\rm cos}(n\varphi)$ and ${\rm sin}\varphi\to{\rm sin}(n\varphi)$.
In what follows we also employ the abbreviations
\begin{equation}
s_i={\rm sin}\xi_i \qquad {\rm and}\qquad 
c_i={\rm cos}\xi_i
\label{eq:abbr}
\end{equation}
for the trigonometrical functions of the isospin angles $\xi_1$ 
and $\xi_2$. A global rotation within the plane of the second 
and third components by an angle 
$\alpha$ with ${\rm tan}\alpha=s_1c_2/c_1$
transforms the unit vector $\hat{\Vek{n}}$ into
\begin{equation}
\widetilde{\Vek{n}}(\xi_1,\xi_2,\varphi)=\begin{pmatrix}
s_1s_2 {\rm cos}\varphi\cr \sqrt{1-s_1^2s_2^2} \cr
0 \cr s_1s_2 {\rm sin}\varphi
\end{pmatrix}\,.
\label{eq:nhat1}
\end{equation}
Hence observables (which are, by definition, gauge invariant) will not 
depend on the two angles $\xi_1$ and $\xi_2$ individually but only on the 
product $s_1s_2$. Stated otherwise, all observables must remain 
invariant along paths of constant $s_1s_2$ in isospin space~\cite{Klinkhamer:1997hw}.

The unit vector $\hat{\Vek{n}} = (n_0,\Vek{n}) \in S^4$ defines the $SU(2)$ matrix
$U(\xi_1,\xi_2,\varphi)=n_0\ID-i\Vek{n}\cdot\Vek{\tau}$, where
$\Vek{\tau}=(\tau^1, \tau^2,\tau^3)$ are the three Pauli matrices.
The Higgs and gauge fields of the string are then characterized 
by two profile functions $f_H$ and $f_G$ that are functions of the
distance ($\rho$) from the string center:
\begin{align}
\begin{pmatrix} \phi_+(\rho,\varphi) \\[1mm] 
\phi_0(\rho,\varphi) \end{pmatrix}=
 f_H(\rho) U(\xi_1,\xi_2,\varphi)
\begin{pmatrix}0 \\[1mm] v \end{pmatrix}
\qquad {\rm and}\qquad
\Vek{W}(\rho,\varphi)= \frac{1}{g}\,
\frac{\hat{\Vek{\varphi}}}{\rho} f_G(\rho)\,
U(\xi_1,\xi_2,\varphi)\,\partial_\varphi
U^\dagger(\xi_1,\xi_2,\varphi)\,.
\label{eq:profiles}\end{align}
Here, the gauge connection $\Vek{W}$ is a vector in coordinate space and 
a matrix in the adjoint representation of weak isospace. The profile functions 
vanish at the core of the string ($\rho=0$) and approach unity at spatial infinity. From this 
parametrization we find the classical mass of the string\footnote{Here and in the following, 
the prime indicates a derivative with respect to the radial argument $\rho$, and we omit 
the argument for simplicity if no confusion can occur.}
\begin{equation}
\frac{E_{\rm cl}}{m^2}=2\pi\int_0^\infty \rho\, d\rho\,\left\{
(s_1s_2)^2\,\biggl[\frac{2}{g^2}
\left(\frac{f_G^\prime}{\rho}\right)^2
+\frac{f_H^2}{f^2\rho^2}\,\left(1-f_G\right)^2\biggr]
+\frac{f_H^{\prime2}}{f^2}
+\frac{\mu_h^2}{4f^2}\left(1-f_H^2\right)^2\right\}\,,
\label{eq:ecl}
\end{equation}
where the dimensionless radial integration variable is related to the physical 
radius by $\rho_{\rm phys}=\rho/m$ and we have introduced the mass ratio 
$\mu_H\equiv m_H/m$. As expected, the classical mass only depends on the isospin 
angles via the combination $s_1s_2$, which reflects gauge invariance.

Note that the configuration, Eq.~(\ref{eq:profiles}) approaches a local gauge 
transformation of the constant vacuum configuration at spatial infinity. As a 
consequence, this configuration is not appropriate for techniques that 
require some kind of perturbative expansions which do not preserve 
gauge invariance order by order. In particular, individual Fourier 
transformations of the Higgs and gauge fields are ill defined. We therefore introduce 
an additional radial function $\xi(\rho)$ with the boundary values $\xi(0)=0$ and 
$\lim_{\rho\to\infty}\xi(\rho)=\xi_1$ to define the local $SU_L(2)$ 
gauge transformation
\begin{equation}
V={\rm exp}\left[-i\Vek{\tau}\cdot\Vek{\xi}(\rho,\varphi)\right]
\qquad {\rm with}\qquad
\Vek{\xi}(\rho,\varphi)=\xi(\rho)\,\begin{pmatrix}
s_2\, {\rm cos}\varphi \cr 
-s_2\, {\rm sin}\varphi \cr c_2
\end{pmatrix}\,.
\label{eq:localgauge}
\end{equation}
Since $\xi(0)=0$ this gauge transformation does not introduce any
singularity at the origin; at spatial infinity it accounts for the above 
mentioned gauge transformation of the constant vacuum. With the gauge 
transformation, Eq.~(\ref{eq:localgauge}) applied, perturbative expansions 
can be performed. Of course, this comes at the expense of an additional
radial function. By construction, observables are independent of its detailed form 
as long as the boundary conditions described above are maintained. For the particular 
case of $\xi_2=\frac{\pi}{2}$ this was verified in Ref.\cite{Weigel:2010pf}. 
In the present study, we will also consider deviations from that particular 
parameter value. We emphasize that the introduction of the gauge rotation, 
Eq.~(\ref{eq:localgauge}) has effectively made our test isospin symmetry 
\emph{local}, since  $\xi_1$ has turned into a space dependent quantity.

To write down the Dirac Hamiltonian from which we compute the spectrum 
of the fermion fluctuations we extract the Hamiltonian, $\mathcal{H}$
from the Lagrangian, Eq.~(\ref{gaugelag}) and then perform the 
left--handed gauge transformation defined in Eq.~(\ref{eq:localgauge}):
$H=\left(P_R+VP_L\right)\mathcal{H}\left(P_R+VP_L\right)^\dagger$. To 
simplify the presentation we define $\Delta(\rho)\equiv\xi_1-\xi(\rho)$ 
and separate the interaction part (again using dimensionless variables)
\begin{eqnarray}
H&=&-i\begin{pmatrix}0 & \Vek{\sigma}\cdot\hat{\Vek{\rho}} \cr
\Vek{\sigma}\cdot\hat{\Vek{\rho}} & 0\end{pmatrix} \partial_\rho
-\frac{i}{\rho}\begin{pmatrix}0 & \Vek{\sigma}\cdot\hat{\Vek{\varphi}}\cr
\Vek{\sigma}\cdot\hat{\Vek{\varphi}} & 0\end{pmatrix} \partial_\varphi
+\begin{pmatrix} 1 & 0 \cr 0 &-1\end{pmatrix}
+H_{\rm int}\,, \label{eqDirac0}\\[4mm]
H_{\rm int}&=&
\left[\left(f_H{\rm cos}(\Delta)-1\right)
\begin{pmatrix} 1 & 0 \cr 0 &-1\end{pmatrix}
+if_H\,{\rm sin}(\Delta)\begin{pmatrix}0 & 1 \cr -1 & 0\end{pmatrix}
I_H\right]
+\frac{1}{2}\frac{\partial \xi}{\partial \rho}
\begin{pmatrix}-\Vek{\sigma}\cdot\hat{\Vek{\rho}}
& \Vek{\sigma}\cdot\hat{\Vek{\rho}} \cr
\Vek{\sigma}\cdot\hat{\Vek{\rho}}
& -\Vek{\sigma}\cdot\hat{\Vek{\rho}}\end{pmatrix}I_H
\nonumber \\[3mm]
&&
+\frac{s_2}{2\rho}\, \begin{pmatrix}
-\Vek{\sigma}\cdot\hat{\Vek{\varphi}}
& \Vek{\sigma}\cdot\hat{\Vek{\varphi}} \cr
\Vek{\sigma}\cdot\hat{\Vek{\varphi}}
& -\Vek{\sigma}\cdot\hat{\Vek{\varphi}}\end{pmatrix}
\Big[f_G\,{\rm sin}(\Delta)\,I_G(\Delta)
+(f_G-1)\,{\rm sin}(\xi)\,I_G(-\xi)\Big]\,.
\label{eqDirac}
\end{eqnarray}
The isopsin matrices in this expression are 
\begin{equation}
I_H=\begin{pmatrix}
c_2 & s_2 {\rm e}^{i\varphi} \cr 
s_2 {\rm e}^{-i\varphi} & -c_2 \end{pmatrix}
\qquad {\rm and} \qquad
I_G(x)=\begin{pmatrix}
-s_2{\rm sin}(x) & [c_2{\rm sin}(x)-i{\rm cos}(x)]\,{\rm e}^{i\varphi} \cr
[c_2{\rm sin}(x)+i{\rm cos}(x)]\,{\rm e}^{-i\varphi} &
s_2{\rm sin}(x)\end{pmatrix}\,.
\label{eq:IG}
\end{equation}
Note that the latter appears with different arguments in Eq.~(\ref{eqDirac}).
Nothing from the invariance along the path with $s_1s_2={\rm const.}$ is
manifest in Eq.~(\ref{eqDirac}), neither is the gauge invariance from 
Eq.~(\ref{eq:localgauge}).

\bigskip\noindent
To proceed, we diagonalize the Hamiltonian in a basis of wave functions
\begin{equation}
\Psi_{\ell}(\rho,\varphi) = \sum_{s,j = \pm \frac{1}{2}}
\Big(\langle \rho\,|\, \langle\,\varphi\,;\,S\,I\,| \Big)\,
|\epsilon \,\ell\,s\,j\,\rangle \,.
\label{eq:GSansatz}
\end{equation}
that decouple radial and angular coordinates in the upper
and lower components of the Dirac spinors 
($\epsilon$ refers to the energy eigenvalue
defined in Eq.~(17) below)
\begin{equation}
\begin{array}{ll}
\langle \rho |\epsilon\,\ell++\rangle =
\begin{pmatrix}f_1(\rho)|\ell + +\rangle \cr
g_1(\rho)|\ell - +\rangle \end{pmatrix} & \qquad
\langle \rho |\epsilon\,\ell+-\rangle =
\begin{pmatrix}f_2(\rho)|\ell + -\rangle \cr
g_2(\rho)|\ell - -\rangle \end{pmatrix}
\cr \cr
\langle \rho |\epsilon\,\ell-+\rangle =
\begin{pmatrix}f_3(\rho)|\ell - +\rangle \cr
g_3(\rho)|\ell + +\rangle \end{pmatrix}& \qquad
\langle \rho |\epsilon\,\ell--\rangle =
\begin{pmatrix}f_4(\rho)|\ell - -\rangle \cr
g_4(\rho)|\ell + -\rangle \end{pmatrix}\,,
\end{array}
\label{eq:GSspinors}
\end{equation}
The notation is such that the signs denote the spin 
and isospin projection quantum numbers. For instance,
\begin{equation}
\langle \varphi;SI|\ell + +\rangle=
{\rm e}^{i(\ell+1)\varphi}\,
\begin{pmatrix}1 \cr 0 \end{pmatrix}_S
\otimes\begin{pmatrix}1 \cr 0 \end{pmatrix}_I  \,.
\label{eq:spinexample}
\end{equation}
Diagonalization means that we construct the eigenvalues
of the stationary Dirac equation
\begin{equation}
H\Psi=\epsilon\Psi\,,
\label{eqDirac1}
\end{equation}
with $|\epsilon|<1$. For $|\epsilon|>1$ we construct the full
scattering matrix as a function of momentum $k=\sqrt{\epsilon^2-1}$. 
Since we employ four component Dirac spinors, we have 
$\left\{H,\alpha_3\right\}=0$ and the spectrum is charge conjugation 
invariant. Our final result of the scattering problem (described in 
appendix A) is the Jost function $\nu(t)$ for imaginary momentum 
$k=it$, as well as the first two terms of its Born expansion obtained by 
iterating the interaction part $H_{\rm int}$.

\section{Vacuum Polarization Energy (VPE)}
\label{sec:vpe}
The main goal of the present investigation is to verify that our 
treatment of the ultraviolet divergences does not produce any 
dependence on the isospin angles $\xi_1$ and $\xi_2$ that cannot be 
expressed as $s_1s_2$. Any change in the renormalization conditions is 
described by finite counterterms. As was the case for the classical energy
Eq.~(\ref{eq:ecl}), the counterterms are manifestly functions of $s_1s_2$. 
We are therefore free to employ the simplest renormalization scheme, which 
is $\overline{\rm MS}$. For the profile functions we choose a specific 
form and introduce dimensionless width parameters $w_G$, $w_H$ and $w_\xi$: 
\begin{equation}
f_H(\rho)=1-\exp\left(-\frac{\rho}{w_H}\right)\,,\quad
f_G(\rho)=1-\exp\left(-\frac{\rho^2}{w_G^2}\right)
\quad \mbox{and} \quad
\xi(\rho)=\xi_1\left[1-\exp\left(-\frac{\rho^2}{w_\xi^2}\right)\right]\,.
\label{eqn:profile}
\end{equation}
Observable values for the width parameters are in units of $m^{-1}$ since
$\rho_{\rm phys}=\rho/m$. Recall again that $\xi(\rho)$ is just an auxiliary
profile describing the local gauge transformation, Eq.~(\ref{eq:localgauge}), 
and that the \emph{VPE} should be independent of $w_\xi$. 
With these conventions on the ansatz parameters, the \emph{VPE}
depends on the model parameters $g$, $f$ and $v$ only via the overall factor
$m^2=(vf)^2$, see also Eq.~(\ref{eq:ecl}). In this sense the dependence 
on the model parameters is completely contained in the classical energy and 
the counterterms, and thus requires little numerical effort.

\medskip\noindent
The spectral method \cite{Graham:2009zz} to compute the \emph{VPE} from scattering
data identifies the change of the density of states caused by a static background 
as the derivative of the scattering phase shift (also known as the phase of the Jost 
function for real momenta)  via the Krein--Friedel--Lloyd formula, 
cf.~Ref.~\cite{Faulkner:1977aa} and references therein. More precisely, we obtain the
phase shift as $(-i/2){\rm ln}({\rm det}S)$, where $S$ is the scattering matrix
of the multichannel problem. Integration over the momentum along the string then
yields the \emph{VPE} per unit length. However, that integral is only finite due to 
particular sum rules among the scattering data \cite{Graham:2001iv}. Ultimately 
this leads to the interface formalism \cite{Graham:2001dy} in which we only need 
to integrate over the momentum $k$ of the scattering problem in the plane perpendicular 
to the string. In this situation, it is prudent to use the analytic properties 
of the scattering data to perform the final momentum integral over imaginary 
momentum $t$ with $k=it$. This analytic continuation has several advantages: First, 
it allows to interchange the momentum integral with the angular momentum 
sum~\cite{Schroder:2007xk} and second, it implicitly collects the contributions 
to the \emph{VPE} coming from the bound states. This is beneficial, as there is 
generally a large number of such states, in particular for wide strings, and
identifying them numerically is cumbersome. To express the \emph{VPE} as an integral over 
imaginary momenta it is essential that the scattering phase shift is an odd function 
of the real momentum. Typically this property results from the Hamiltonian being 
real~\cite{Chadan:1977pq,Newton:1982qc} which is, however, not the case here: 
The gauge transformation, Eq.~(\ref{eq:localgauge}) turns the global isospin transformation
along the path $s_1 s_2={\rm const.}$ into a local one 
and, consequently, there is
no global transformation on the basis states, Eq.~(\ref{eq:GSspinors}) which could result in
a real Hamiltonian.\footnote{The Hamiltonian is still Hermitian, of course, and the 
single particle energies are real.} In Appendix~A we show that nevertheless the phase 
shift is odd in the momentum. 

After collecting all information the \emph{VPE} per unit length of the string is expressed as 
\begin{equation}
E_{\rm vac}=\frac{m^2}{2\pi} \int_0^\infty d\tau\, \tau
\left\{\sum_{\ell} D_{\ell}\left[\nu(\tau,\ell)-\nu_1(\tau,\ell)-\nu_2(\tau,\ell)\right]
-\frac{c_F}{c_B}\sum_{\ell}\bar{D}_\ell\bar{\nu}_2(\tau,\ell)\right\}
+E_2+E_{\rm f.b.}\,,
\label{eq:master}
\end{equation}
where we performed a final change of variable $t\to\tau=\sqrt{t^2-1}$ to avoid the 
integrable singularity at $t=m$. In Eq.~(\ref{eq:master}) $\nu$ is the full Jost function 
with orbital angular momentum $\ell$ and degeneracy factor $D_\ell=2-\delta_{\ell,-1}$ 
on the imaginary momentum axis, while $\nu_1$ and $\nu_2$ are first two terms of its
Born expansion with respect to $H_{\rm int}$. These two subtractions are performed 
before summing over angular momentum channels. This is indispensable in order to 
identify and disentangle the subleading logarithmic divergence and the relevant finite 
contributions from the two leading Born terms. In fact, the logarithmic divergence has 
additional contributions from the third and fourth order Feynman diagrams, and their 
total strength\footnote{Here, the term ``strength'' means that the Feynman diagrams
produce the singularity $\frac{c_F/2\pi}{4-D}$ in dimensional regularization. 
In Ref.~\cite{Graham:2011fw} a factor 4 was omitted in the definition of both $c_F$ 
and $c_B$, so that the ratio remains unaffected.}
is $c_F$. The second order contribution of quantum corrections from a complex boson field 
about a static background also produces a logarithmic divergence. Let $c_B$ be its 
strength and $\bar{\nu}_2(\tau,\ell)$ the second order Born term of its Jost function for 
imaginary momenta in the angular momentum channel $\ell$. Then the last term 
in curly brackets of Eq.~(\ref{eq:master}) removes the logarithmic divergence 
from the integral. Since there is no further (sub--subleading)
divergence, this subtraction can be made after summing over angular 
momenta. In the last step all subtractions are added back in the form of 
Feynman diagrams. They are computed by standard techniques using, {\it e.g.},
dimensional regularization. Their divergent parts are uniquely compensated 
by counterterms in a definite renormalization scheme. All that remains 
are the finite parts $E_2$ and $E_{\rm f.b.}$ of the second order fermion and 
fake boson diagrams, which correspond to the finite parts of the subtractions 
$\nu_{1,2}$ and $\bar{\nu}$, respectively. Equation~(\ref{eq:master}) is the master 
formula to compute the \emph{VEV} of string configurations.

We stress that only the very first term under the integral in eq.~(\ref{eq:master}) 
remains unchanged when varying the string isospin orientation, provided that
${\rm sin}(\xi_1){\rm sin}(\xi_2)$ remains constant. All other contributions are 
more general functions of $\xi_1$ and $\xi_2$ and thus vary along our particular isospin path of 
constant ${\rm sin}(\xi_1){\rm sin}(\xi_2)$. 
These terms should eventually cancel provided that the 
identity of Born and Feynman series holds. However, individually they
represent ill defined ultraviolet divergent quantities that 
undergo distinct regularization procedures and it is therefore unclear whether the 
spectral approach and, in particular, the renormalization procedure spoil 
gauge invariance. We will investigate this question numerically in the next section.

\section{Results}

\begin{figure}
\centerline{
\epsfig{file=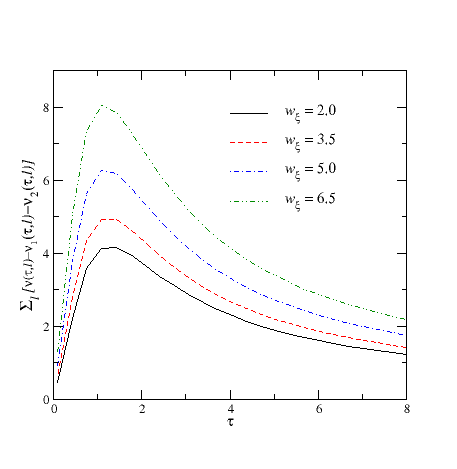,width=8cm,height=7.0cm}\hspace{1cm}
\epsfig{file=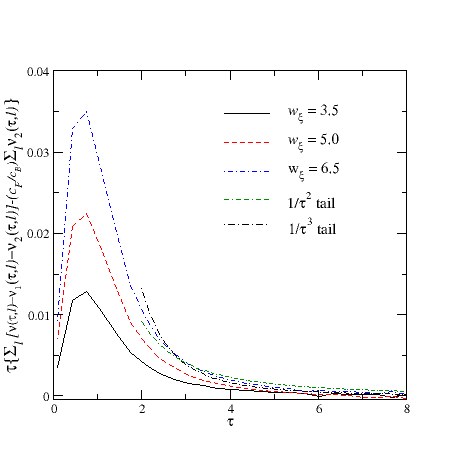,width=8cm,height=7.0cm}}
\caption{\label{fig:diffwx}(Color online) Partial sums that enter the \emph{VPE}, 
Eq.~(\ref{eq:master}), for the string profiles with $w_G=w_H=4.82$, $\xi_1=0.3\pi$ 
and $\xi_2=0.25\pi$. The left panel shows the fermion part for four different 
values of $w_\xi$. The right panel shows the total integrand for three values 
of $w_\xi$ relative to $w_\xi=2.0$. The double--dashed lines that start 
at $\tau=2$ in the right panel are simple power decays which serve to 
guide the eye on the large $\tau$ behavior. Note the different 
scales in the two panels.}
\end{figure}

The computation of the momentum integral and its integrand in Eq.~(\ref{eq:master}) 
is by far the most expensive part of the numerical procedure. To begin with, 
the $\ell=-1$ and $\ell=0$ channels require particular consideration. They 
involve Hankel functions of order zero whose irregular component diverges 
logarithmically at small arguments rather than by an inverse power law. Thus 
regular and irregular components are numerically difficult to separate. When 
integrating the radial differential equation [Eqs.~(\ref{deqGFin2}) and~(\ref{defDZud}) 
for $k=it$] we take the lower boundary to be $\rho_{\rm min}\sim 10^{-50}$ for these 
two channels, and from $\rho_{\rm min}$ we  extrapolate to 
$\rho=0$. In other channels a lower boundary of $\rho_{\rm min}\sim 10^{-12}$ 
is fully reliable. Angular momenta are typically summed up to $\ell_{\rm max}=600$ 
or $\ell_{\rm max}=700$ above which numerical stability for Hankel functions at 
small arguments is lost. For background profiles with small or moderate widths this 
gives sufficient accuracy. Once the angular momentum sum is completed, the analog 
contribution from the fake boson (mimicking the logarithmic ultraviolet divergences 
from third and fourth order Feynman diagrams) is subtracted and the large $\tau$ 
behavior of the integrand is treated by fitting a $1/\tau^3$ tail, {\it cf.} 
the right panel of Fig.~\ref{fig:diffwx}. Finally, 
for wider profiles an additional extrapolation of the angular momentum sum to
$\ell_{\rm max}\to\infty$ is necessary which typically adds about $1\ldots2 \%$ 
to the \emph{VPE}.

We start with a few examples, displayed in Table~\ref{tab:wx} and Fig. 
\ref{fig:diffwx}, in order to verify the independence from the gauge 
profile $\xi(\rho)$.
\begin{table}
\centerline{
\begin{tabular}{c|c|c|c||c||c}
$w_\xi$ & $E_{\delta}$ & $c_F$ &  $E_{\rm FD}$ & $E_{\rm vac}$
& $|E_{\delta}|+|E_{\rm FD}|$\cr
\hline
2.0 & 0.3010 & -10.00 & -0.0108 &~~0.2902~~~& 0.3118 \cr
3.5 & 0.2974 & -11.59 & -0.0072 &~~0.2902~~~& 0.3046 \cr
5.0 & 0.2953 & -14.29 & -0.0047 &~~0.2905~~~& 0.3000 \cr
6.5 & 0.2915&  -17.82 & -0.0015 &~~0.2901~~~& 0.2930
\end{tabular}}
\caption{\label{tab:wx}Example for the invariance with respect to
the local gauge transformation, Eq.~(\ref{eq:localgauge}) with
$E_{\rm FD}=E_2+E_{\rm f.b.}$. Listed are all ingredients from 
Eq.~(\ref{eq:master}) that explicitly depend on
the width $w_\xi$ of the gauge profile $\xi(\rho)$.
Parameters are $w_G=w_H=4.82$, $\xi_1=0.3\pi$ and $\xi_2=0.25\pi$.
In all cases an identical fake boson profile was employed because
it affects $E_{\rm FD}$.}
\end{table}
The variation of the individual contributions to the \emph{VPE} is an order of 
magnitude larger than that of the total result. The tiny variation of the 
latter is due to errors from the numerical simulation. The cancellation 
of the gauge variant parts for the \emph{VPE} is most obvious when adding them 
as absolute values which contains spreads of up to 10\%. A large variation 
appears in the fermion part of the momentum integral (\emph{i.e.}~the 
contribution from the first term in curly brackets) in Eq.~(\ref{eq:master}),
as can be seen in Fig.~\ref{fig:diffwx}. 

Even though we have just established that the \emph{VPE} does not vary 
with the width of the gauge profile, it is prudent for numerical efficiency 
and stability to choose that width similar to  one in the profile 
functions of the physical boson fields, because otherwise large angular momenta 
play too significant a role.

\begin{table}
\centerline{
\begin{tabular}{c|c|c|c|c||c||c}
$\xi_1/\pi$ &$\xi_2/\pi$ & $E_{\delta}$ & $c_F$ &  $E_{\rm FD}$ & $E_{\rm vac}$
& $|E_{\delta}|+|E_{\rm FD}|$\cr
\hline
0.1 & 0.4 & 0.1504 & -4.913 &  0.0014 &~~0.1518~~~& 0.1518\cr
0.4 & 0.1 & 0.1702 & -8.541 & -0.0180 &~~0.1521~~~& 0.1882 \cr
0.3 & 0.11834 & 0.1496 & -6.814 & 0.0021 &~~0.1517~~~& 0.1517 \cr
0.2 & 1/6 & 0.1639 & -5.615 & -0.0117 &~~0.1522~~~& 0.1758
\end{tabular}}
\caption{\label{tab:zz} Contributions to Eq.~(\ref{eq:master}) and their 
variations with the isospin angles. In all cases we have $s_1s_2\approx0.29389$. 
The width parameters of the boson profiles are $w_G=w_H=3.5$. The results were
obtained with various values for the widths of the gauge and fake boson
profiles.}
\end{table}

\begin{figure}
\centerline{
\epsfig{file=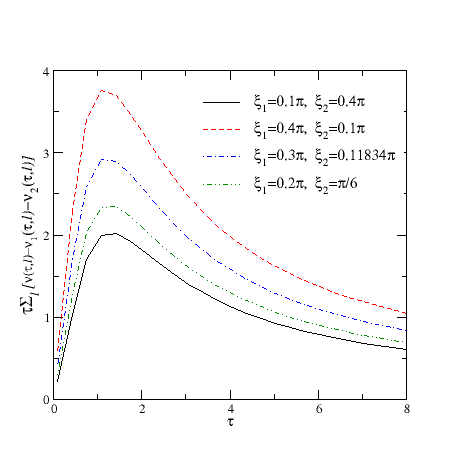,width=9cm,height=6cm}}
\caption{\label{fig:zz}(Color online) The fermion part of the momentum integrand in 
Eq.~(\ref{eq:master}) (similar to the left panel of Fig.~\ref{fig:diffwx}).
The selected width parameters are $w_H=w_G=3.5$.}
\end{figure}
In Fig.~\ref{fig:zz} we show the strongly varying fermion part of the integrand 
for the \emph{VPE} for sets of isospin angles that produce identical products 
$s_1s_2$. Despite the pronounced variation of this particular piece, the total 
\emph{VPE} only differs at the order of the numerical accuracy as can clearly be seen 
from the data in Table~\ref{tab:zz}. The comparison with the (incorrect) addition 
of the absolute values of the gauge variant contributions further illustrates this 
observation.

\begin{figure}
\centerline{
\epsfig{file=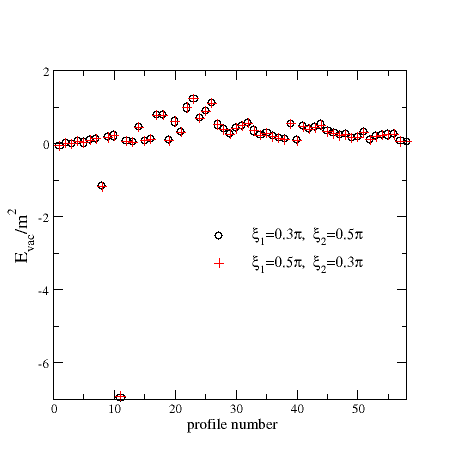,width=8cm,height=6.5cm}\hspace{1cm}
\epsfig{file=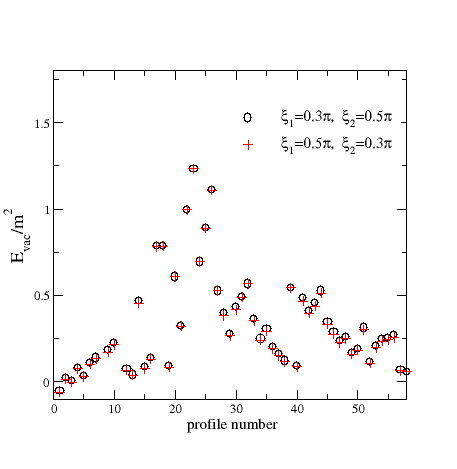,width=8cm,height=6.5cm}}
\caption{\label{fig:x1x2A}(Color online) The vacuum polarization energy for different 
background profiles with the two isospin angles swapped. In the right panel we zoom 
in by omitting narrow profiles that suffer from the Landau ghost 
problem \cite{Ripka:1987ne,Hartmann:1994ai,Graham:2011fw}. Details of the profiles 
are listed in appendix B.}
\end{figure}

\begin{figure}
\centerline{
\epsfig{file=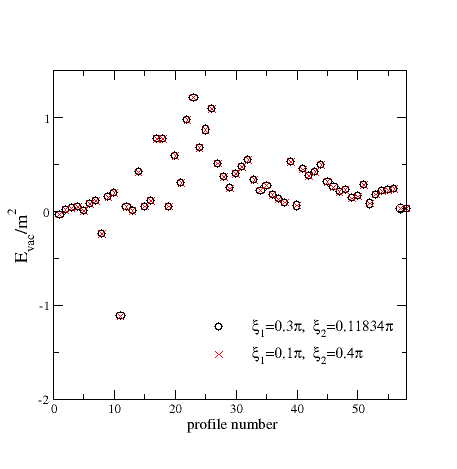,width=8cm,height=6.5cm}\hspace{1cm}
\epsfig{file=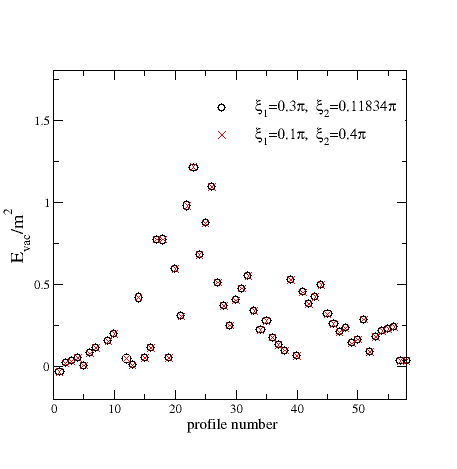,width=8cm,height=6.5cm}}
\caption{\label{fig:x1x2B}Same as Fig.~\ref{fig:x1x2A} for 
a second pair of isospin angles.}
\end{figure}

For  $\xi_2=\frac{\pi}{2}$ the Hamiltonian is real. In this 
simpler case the \emph{VPE} was computed for about 50 sets of width parameters 
($w_H$, $w_G$, {\it cf.} Appendix B) and eight different values for 
$\xi_1\in[0,\frac{\pi}{2}]$ in Ref.~\cite{Graham:2011fw}. These results\footnote{We 
have reproduced these earlier results for $\xi_2=\frac{\pi}{2}$ 
using the more general numerical simulation for the complex Hamiltonian.} were then 
used to establish stable charged cosmic strings for fermion masses only slightly
larger than that of the top quark. Here we consider the same sets of width parameters
for two pairs of isospin angles that yield the identical products $s_1s_2$. In the 
first of the two pairs we simply swap the isospin angles as compared to the earlier 
calculations \cite{Graham:2011fw}, and show the resulting 
\emph{VPE} (in the $\overline{\rm MS}$ renormalization scheme) in Fig.~\ref{fig:x1x2A}.
Obviously the computed \emph{VPE}s agree within the numerical accuracy for the full 
range of considered width parameters. However, merely swapping the isospin angles is 
not sufficient to fully establish dependence on only the product $s_1s_2$. For 
example, there could be gauge variant contributions involving ${\rm sin}(\xi_1+\xi_2)$. 
To rule out such a dependence, we have made a second study and
compared the two sets $(\xi_1,\xi_2)=(0.1,0.4)\pi$ and $(\xi_1,\xi_2)=(0.3,0.11834)\pi$. 
The resulting \emph{VPEs} are shown in Fig.~\ref{fig:x1x2B}.
Again we observe perfect agreement for the computed \emph{VPEs} as the tiny numerically 
discrepancies are not resolved within Figs.~\ref{fig:x1x2A} and \ref{fig:x1x2B}.
So we conclude that the spectral methods to compute the \emph{VPE} of cosmic strings 
indeed preserve gauge and isospin invariance even though some of its components do not.

The comparison of the results in Fig.~\ref{fig:x1x2A} with those in Fig.~\ref{fig:x1x2B} 
suggests that the \emph{VPE} depends on the isospin orientation only mildly, 
except for the very narrow configurations that suffer from the Landau ghost 
problem~\cite{Ripka:1987ne,Hartmann:1994ai,Graham:2011fw}. This is not quite the case: 
in the current study our goal is to compare the \emph{VPE} for configurations with equal 
$s_1s_2$, as in either of Figs.~\ref{fig:x1x2A} or \ref{fig:x1x2B}.
To reveal the discussed invariance, the difference between the two angles 
is usually chosen deliberately large, so that
one of the angles is always small and so is the product $s_1s_2$. 
When we lift this restriction we find e.g.~with $w_G=w_H=6.0$ that 
$E_{\rm vac}$ increases from $0.438m^2$ to $0.479m^2$ between 
$s_1s_2=0$ and $s_1s_2=1$.

In a separate study we have implemented a boundary condition at large 
separation from the string to construct discretized basis states 
that serve to compute matrix elements of the Dirac Hamiltonian, Eq.~(\ref{eqDirac}). 
These matrix elements form a complex Hermitian matrix that we have diagonalized 
using \texttt{LAPACK}~\cite{laug1999}. Eigenvalues below threshold are identified 
as bound state energies. We have verified that all energy eigenvalues of the Dirac 
Hamiltonian remain unchanged when altering $\xi_1$ and $\xi_2$ such that $s_1s_2$ 
stays constant. This is expected for bound states that have no support in the 
vicinity of the boundary. Scattering states, however, reach out to spatial 
infinity and are thus sensitive to the discretizing boundary conditions which 
are not manifestly gauge invariant; so the invariance of these states
comes as some surprise. In addition, this discretization approach 
requires to impose a numerical cutoff on the energy to produce a finite 
dimensional Hamiltonian matrix. The levels slightly below that cutoff 
exhibit a soft variation along the path of invariance in isospace. 
This reflects the fact that unitarity of the transformation is lost for a 
finite dimensional Hilbert space. Similarly, such near-cutoff energies 
do also vary with the gauge profile $\xi(\rho)$. Renormalized 
\emph{VPE} calculations based on this or similar numerical discretization 
approaches \cite{Diakonov:1993ru} will probably be erroneous.
In the spectral approach, we consequently use the discretization technique 
only for the bound states, while scattering states are treated 
in the continuum formulation.

Finally we note in passing that we have numerically verified the bound state energies 
from the above discretization computation against the roots of the Jost function on 
the imaginary axis and also ensured that the number of bound states satisfies 
Levinson's theorem.\footnote{For the bound states the discretization 
procedure is advantageous because root finding algorithms may fail to identify degenerate 
bound states that appear in multi--channel scattering. Also identifying the roots very  
close to threshold is numerically cumbersome.}

\section{Conclusion}

There are numerous obstacles in computing the \emph{VPE} 
of string type configuration in gauge theories that are similar to the standard 
model of particle physics. Within the so--called spectral approach, these obstacles 
can be overcome by an interplay of techniques which individually are \emph{not} 
gauge invariant. If the spectral approach is a meaningful tool in gauge theories, 
it must ensure that the gauge variant contributions eventually cancel. To the best
of our knowledge there is no formal proof of this cancellation at the moment, and it
is also far from obvious  because the gauge-variant contributions are related to 
ultraviolet divergent quantities that undergo different methods of regularization. 
Hence analytical or numerical verifications of gauge invariance in the 
spectral approach are indispensable.

In the present study we have therefore comprehensively revisited the 
computation of the \emph{VPE} for string type configurations arising from 
fermion fluctuations, in order to justify and validate earlier computations 
(carried out in a limited parameter space) that suggested novel solutions 
in theories closely related to the standard model \cite{Weigel:2010zk}. Those 
earlier studies were implicitly based on the assumption that the spectral method 
would not spoil gauge invariance as the identification of Born and Feynman series 
would hold even for (differently regularized) divergent contributions.
Here we have extended the parameter space for an independent numerical corroboration
of this assumption. It employs the invariance of the spectrum of the 
Dirac Hamiltonian along a particular path in the enlarged parameter space. 
This invariance must be reflected in the \emph{VPE}. However, this is not manifest 
in the actual \emph{VPE} calculation, because regularization and renormalization 
indeed require delicate operations on divergent contributions that vary under the 
isospin transformation. 

Our numerical simulations show that individual contributions that are not gauge invariant 
but need to be included for regularization and renormalization may vary by 10\% or more 
along the path of isospin invariance. But then, the contributions combine such that these 
variations actually do cancel in the total result, leading to changes of the fermion 
quantum energy of the cosmic string along the path of isospin invariance of the order of 
only a fraction of a percent. Such variations are within the bounds of the numerical accuracy. 
Thus we have verified numerically that the spectral method preserves gauge invariance and 
is hence a valid tool to study quantum corrections to extended configurations, such as 
cosmic strings in the standard model of particles.

\acknowledgments
H.~W.\ is supported in part by the NRF (South Africa) by Grant No.~77454.  
N.~G.\ is supported in part by the NSF through Grant No.~PHY15-20293.

\appendix

\section{Scattering problem}

Scattering data are essential to the spectral method to compute the \emph{VPE}
because they determine the density of states. After continuing to complex 
momenta, the Jost function on the imaginary axis is the major ingredient.
However, our scattering problem is more general than that typically 
discussed in textbooks \cite{Chadan:1977pq,Newton:1982qc} as the potential is not
real and thus complex conjugation does not produce the second independent
solution. In this appendix we describe the resulting changes up to 
the point where we observe that the sum of the eigenphase shifts
is antisymmetric when reflecting the real momentum $k\to-k$. 
From there, the techniques of Ref. \cite{Graham:2011fw} can be 
copied.

Let $(\vec{f})_j$ and $(\vec{g})_j$ with $j=1,\ldots,4$ denote the linearly 
independent solution of the Dirac equation and  combine them to matrices
\begin{equation}
\begin{array}{r@{\,\,\,\,\longrightarrow\,\,\,\,}ll@{\qquad\mbox{and}\qquad}ll}
(\vec{f})_j & & \displaystyle\left[\mathcal{F} \cdot \mathcal{H}_u\right]_j &
& \displaystyle\left[\mathcal{F}^* \cdot \mathcal{H}^\ast_u\right]_j
\\[3mm]
(\vec{g})_j &  \kappa&\displaystyle\left[\mathcal{G} \cdot \mathcal{H}_d\right]_j &
\kappa& \displaystyle\left[\mathcal{G}^* \cdot \mathcal{H}^\ast_d\right]_j \,.
\end{array}
\label{Smat1}
\end{equation}
in which the free solutions with outgoing boundary conditions 
(recall the we consider unit winding of the string)
\begin{eqnarray}
\mathcal{H}_u&=&\mbox{diag}\left(
H^{(1)}_{\ell+1}(k\rho),H^{(1)}_{\ell}(k\rho),
H^{(1)}_{\ell+2}(k\rho),H^{(1)}_{\ell+1}(k\rho)\right) \\[2mm]
\mathcal{H}_d&=&\mbox{diag}\left(
H^{(1)}_{\ell+2}(k\rho),H^{(1)}_{\ell+1}(k\rho),
H^{(1)}_{\ell+1}(k\rho),H^{(1)}_{\ell}(k\rho)\right)\,,
\label{Smat2}
\end{eqnarray}
have been factorized. The $H^{(1)}_\ell(z)$ are Hankel functions of the first kind and 
describe the outgoing waves. The relative weight of upper and lower Dirac 
components 
\begin{equation}
\kappa \equiv \frac{k}{\epsilon+m} = \frac{\epsilon-m}{k}\,,
\end{equation}
has been introduced to make Hermiticity in the coupled equations
explicit, see below. It is convenient to define $2\times2$ submatrices 
\begin{eqnarray}
H &=& \alpha_H\,\begin{pmatrix} 1 & 0 \cr 0 & 1 \end{pmatrix}\,, \qquad\qquad\qquad
P = \alpha_P\, \begin{pmatrix} -ic_2 & -s_2 \cr s_2 & ic_2 \end{pmatrix}=-P^\dagger\,,
\nonumber \\[3mm]
G &=& \alpha_G\, \begin{pmatrix} 
s_2s_\Delta & c_\Delta+ic_2s_\Delta \cr
c_\Delta-ic_2s_\Delta & -s_2s_\Delta 
\end{pmatrix}
+\alpha_\xi\,\begin{pmatrix} 
-s_2s_\xi & c_\xi-ic_2s_\xi \cr
c_\xi+ic_2s_\xi & s_2s_\xi \end{pmatrix}
+\alpha_r\, \begin{pmatrix} -ic_2 & -s_2 \cr s_2 & ic_2 \end{pmatrix}
\label{defHGpmP}
\end{eqnarray}
Note that for $c_2=0$ and $s_2=1$ these are the matrices as defined
in eq.~(B3) of ref.\cite{Graham:2011fw} with $G_{+}=G$ and 
$G_{-}=G^\dagger$. With these definitions the potential matrices 
become very compact:
\begin{equation}
V_{uu}=\begin{pmatrix} H & G \cr G^\dagger & H \end{pmatrix}\,,
\qquad
V_{dd}=\begin{pmatrix} -H & G^\dagger \cr G & -H \end{pmatrix}\,,
\qquad
V_{ud}=-\begin{pmatrix} G & P \cr P & G^\dagger \end{pmatrix}\,,
\qquad
V_{du}=\begin{pmatrix} -G^\dagger & P \cr P & -G \end{pmatrix}
=V_{ud}^\dagger\,.
\label{eq:compactmatrix}
\end{equation}
Even though the problem is manifestly Hermitian, the matrix elements
are no longer real.

The differential equations for outgoing boundary conditions are 
also discussed in appendix B of ref.\cite{Graham:2011fw}
\begin{eqnarray}
\partial_\rho \mathcal{F} &=&
\left[\overline{\mathcal{M}}_{ff}+O_d\right]\cdot\mathcal{F}
+\mathcal{F}\cdot\mathcal{M}_{ff}^{(r)}
+k\left[\overline{\mathcal{M}}_{fg}+C\right]\cdot\mathcal{G}\cdot Z_d
\cr\cr
\partial_\rho \mathcal{G} &=&
\left[\overline{\mathcal{M}}_{gg}+O_u\right]\cdot\mathcal{G}
+\mathcal{G}\cdot\mathcal{M}_{gg}^{(r)}
+k\left[\overline{\mathcal{M}}_{gf}-C\right]\cdot\mathcal{F}\cdot Z_u\,,
\label{deqGFout}
\end{eqnarray}
where the $4 \times 4$ coefficient matrices without an overline
are purely kinematic,
\begin{align}
Z_u&={\rm diag}\, \left(
\frac{H^{(1)}_{\ell+1}(k\rho)}{H^{(1)}_{\ell+2}(k\rho)}\,,
\frac{H^{(1)}_{\ell}(k\rho)}{H^{(1)}_{\ell+1}(k\rho)}\,,
\frac{H^{(1)}_{\ell+2}(k\rho)}{H^{(1)}_{\ell+1}(k\rho)}\,,
\frac{H^{(1)}_{\ell+1}(k\rho)}{H^{(1)}_{\ell}(k\rho)}\right)
\,, \qquad& Z_d & =\left(Z_u\right)^{-1}\,,
\label{defDZud} \\[4mm]
O_u&=\frac{1}{\rho}\,{\rm diag}\,
\left(-(\ell+2),-(\ell+1),\ell+1,\ell\right)\,, \qquad& 
O_d &=\frac{1}{\rho}\,{\rm diag}\,
\left(\ell+1,\ell,-(\ell+2),-(\ell+1)\right)
\nonumber
\end{align}
and $C = \mathrm{diag}(-1,-1,1,1)$.
The matrices multiplying $\mathcal{F}$ and $\mathcal{G}$ from the right
are also independent of the background potential,
\begin{equation}
\mathcal{M}_{ff}^{(r)}=\mathcal{M}_{ff}^{(r)}(k)=-kC\cdot Z_d(k)-O_d
\qquad {\rm and} \qquad
\mathcal{M}_{gg}^{(r)}=\mathcal{M}_{gg}^{(r)}(k)=kC\cdot Z_u(k)-O_u\,.
\label{kinmatrix}
\end{equation}
Genuine interactions from the string background are solely contained in
the overlined matrices in Eq.~(\ref{deqGFout}). Using the same {$2\times2$}
matrix notation as above, we have explicitly
\begin{equation}
\begin{array}{ll}
\overline{\mathcal{M}}_{gg}=CV_{ud}
=\begin{pmatrix}
G & P \cr -P & -G^\dagger
\end{pmatrix}  & \qquad
\overline{\mathcal{M}}_{ff}=-CV_{du}
=\begin{pmatrix}
-G^\dagger & P \cr  -P & G
\end{pmatrix} \cr \cr
\overline{\mathcal{M}}_{gf}=\frac{1}{E-m}\,CV_{uu}
=\frac{1}{E-m}\begin{pmatrix}
-H & -G \cr  G^\dagger & H
\end{pmatrix}& \qquad
\overline{\mathcal{M}}_{fg}=-\frac{1}{E+m}\,C V_{dd}
=\frac{1}{E+m}\begin{pmatrix}
-H & G^\dagger \cr  -G & H
\end{pmatrix} \,.
\end{array}
\label{intmatrix}
\end{equation}
Note that, in comparison to Ref.\cite{Graham:2011fw}, a factor of $k$ has been 
reshuffled $k$ from the definitions of $\overline{\mathcal{M}}_{gf}$ and
$\overline{\mathcal{M}}_{fg}$ into the differential equations to make the $k$ 
dependence more transparent. Recall also that the factor $k$ [more precisely the
factor $\kappa=k/(E+m)$] arises from the relative weight of the upper and lower
components. Since $E=\sqrt{k^2+m^2}$ the new definitions 
in eq.~(\ref{intmatrix}) are now invariant under $k\leftrightarrow-k$.
The solutions to the differential equations~(\ref{deqGFout}) are subject to
the boundary conditions $\mathcal{F} \to \ID$ and
$\mathcal{G} \to \ID$ at $\rho \to \infty$. 

If the interactions were real the scattering solution and the scattering matrix 
would be defined via Eqs.~(27)--(29) of Ref.\cite{Graham:2011fw}; however, they 
are not. We therefore have to reconstruct the solutions with incoming boundary
conditions explicitly. To this end we introduce (recall that
$H^{(2)}_\nu(x)=\left[H^{(1)}_\nu(x)\right]^\ast$ for real $x$)
\begin{equation}
\overline{Z}_u=\overline{Z}_u(k)={\rm diag}\, \left(
\frac{H^{(2)}_{\ell+1}(k\rho)}{H^{(2)}_{\ell+2}(k\rho)}\,,
\frac{H^{(2)}_{\ell}(k\rho)}{H^{(2)}_{\ell+1}(k\rho)}\,,
\frac{H^{(2)}_{\ell+2}(k\rho)}{H^{(2)}_{\ell+1}(k\rho)}\,,
\frac{H^{(2)}_{\ell+1}(k\rho)}{H^{(2)}_{\ell}(k\rho)}\right)
\qquad {\rm and} \qquad 
\overline{Z}_d=\left(\overline{Z}_u\right)^{-1}
\label{zin}
\end{equation}
that enter
\begin{eqnarray}
\partial_\rho \overline{\mathcal{F}}&=&
\left[\overline{\mathcal{M}}_{ff}+O_d\right]\cdot\overline{\mathcal{F}}
+\overline{\mathcal{F}}\cdot\mathcal{N}_{ff}^{(r)}
+k\left[\overline{\mathcal{M}}_{fg}+C\right]\cdot\overline{\mathcal{G}}\cdot \overline{Z_d}
\cr\cr
\partial_\rho \overline{\mathcal{G}}&=&
\left[\overline{\mathcal{M}}_{gg}+O_u\right]\cdot\overline{\mathcal{G}}
+\overline{\mathcal{G}}\cdot\mathcal{N}_{gg}^{(r)}
+k\left[\overline{\mathcal{M}}_{gf}-C\right]\cdot\overline{\mathcal{F}}\cdot \overline{Z_u}\,,
\label{deqGFin}
\end{eqnarray}
with the additional definitions (note the overline '$\overline{\quad\vphantom{Z}}$' added to
$Z_u$ and $Z_d$ )
\begin{equation}
\mathcal{N}_{ff}^{(r)}=\mathcal{N}_{ff}^{(r)}(k)=-kC\cdot \overline{Z}_d(k)-O_d
\qquad {\rm and} \qquad
\mathcal{N}_{gg}^{(r)}=\mathcal{N}_{gg}^{(r)}(k)=kC\cdot \overline{Z}_u(k)-O_u\,.
\label{kinmatrix2}
\end{equation}
According to Eq.~(9.1.39) in Ref.\cite{AS} we have
$$
H_\nu^{(2)}(z)=-{\rm e}^{i\nu\pi}H_\nu^{(1)}(-z)
=-(-1)^\nu H_\nu^{(1)}(-z)
$$ 
and thus 
$$
\overline{Z_u}(k)=-Z_u(-k)
\qquad {\rm and} \qquad
\overline{Z_d}(k)=-Z_d(-k)\,.
$$
This implies 
$$
\mathcal{N}_{ff}^{(r)}(k)=
\mathcal{M}_{ff}^{(r)}(-k)
\qquad {\rm and} \qquad
\mathcal{N}_{gg}^{(r)}(k)=
\mathcal{M}_{gg}^{(r)}(-k)\,.
$$
Hence the wave equations~(\ref{deqGFin}) can be written as
\begin{eqnarray}
\partial_\rho \overline{\mathcal{F}}&=&
\left[\overline{\mathcal{M}}_{ff}+O_d\right]\cdot\overline{\mathcal{F}}
+\overline{\mathcal{F}}\cdot\mathcal{M}_{ff}^{(r)}(-k)
-k\left[\overline{\mathcal{M}}_{fg}+C\right]\cdot\overline{\mathcal{G}}\cdot Z_d(-k)
\cr\cr
\partial_\rho \overline{\mathcal{G}}&=&
\left[\overline{\mathcal{M}}_{gg}+O_u\right]\cdot\overline{\mathcal{G}}
+\overline{\mathcal{G}}\cdot\mathcal{M}_{gg}^{(r)}(-k)
-k\left[\overline{\mathcal{M}}_{gf}-C\right]\cdot\overline{\mathcal{F}}\cdot Z_u(-k)\,.
\label{deqGFin2}
\end{eqnarray}
Equations~(\ref{deqGFin2}) are also obtained from Eqs.~(\ref{deqGFout}) by
replacing $k\to-k$. Since $\mathcal{F}$, $\mathcal{G}$, $\overline{\mathcal{F}}$ 
and $\overline{\mathcal{G}}$ all obey the same boundary conditions at 
spatial infinity, this implies that
\begin{equation}
\overline{\mathcal{F}}(k)=\mathcal{F}(-k)
\qquad {\rm and} \qquad
\overline{\mathcal{G}}(k)=\mathcal{G}(-k)\,.
\label{signk}
\end{equation}
The scattering solution constructed from the $\mathcal{F}$ components read
\begin{equation}
\Psi = \overline{\mathcal{F}}\cdot \mathcal{H}_u^\ast -
(\mathcal{F}\cdot \mathcal{H}_u) \cdot \mathcal{S}
\label{scatsol}
\end{equation}
and regularity at $\rho\to0$ determines the scattering matrix
\begin{equation}
\mathcal{S}=\lim_{\rho\to0}\,
\mathcal{H}_u^{-1}\cdot\mathcal{F}^{-1}\cdot
\overline{\mathcal{F}}\cdot\mathcal{H}_u^\ast\,.
\label{Smat3}
\end{equation}
The sum of the eigenphase shifts thus finally is
\begin{equation}
\delta_\ell(k)=\frac{1}{2i}\,{\rm ln} \,{\rm det}\,
\lim_{\rho\to0}\mathcal{F}_\ell(\rho,k)^{-1}\cdot
\overline{\mathcal{F}}_\ell(\rho,k)
=\frac{i}{2}\left[{\rm det} \,{\rm tr}\,
\lim_{\rho\to0}\mathcal{F}_\ell(\rho,k)
-{\rm det} \,{\rm tr}\,
\lim_{\rho\to0}\mathcal{F}_\ell(\rho,-k)\right]\,,
\label{delta}
\end{equation}
where we have restored all the arguments and made use of the reflection
symmetry derived in Eq.~(\ref{signk}). This clearly shows that the
eigenphase shift is odd under $k$. Thus the phase shift part of the
VPE can indeed be computed from the Jost function at imaginary 
momenta \cite{Graham:2009zz}. In Ref.~\cite{Graham:2011fw}
the derivation of the entries in Eq.~(\ref{eq:master}) from
continuation of Eqs.~(\ref{deqGFout}) or~(\ref{deqGFin}) has been
discussed in detail and must not be repeated here.

\section{Radial parameters}

In this appendix we list, within Table~\ref{tab:para}, the details of the 
background profiles that were used for the numerical simulations in Sec.~IV. 
The definition of the variational width parameters is given in Eq.~(\ref{eqn:profile}).
\begin{table}
\begin{tabular}{c|c|c||c|c|c||c|c|c||c|c|c||c|c|c}
$n$ & $w_H$ & $w_G$ &  $n$ & $w_H$ & $w_G$ 
& $n$ & $w_H$ & $w_G$ & $n$ & $w_H$ & $w_G$& $n$ & $w_H$ & $w_G$\cr
\hline
1 & 0.5 & 0.5 & 13& 1.0 & 2.0 & 25 & 8.5 & 8.5 & 37 & 3.25 & 3.25 & 49 & 3.35 & 3.35 \cr
2 & 0.5 & 2.0 & 14& 6.0 & 6.0 & 26 & 9.5 & 9.5 & 38 & 2.75 & 2.75 & 50 & 3.62 & 3.62 \cr
3 & 2.0 & 0.5 & 15& 1.0 & 3.0 & 27 & 6.5 & 6.5 & 39 & 6.6 & 6.6 & 51 & 4.82 & 4.82 \cr
4 & 2.0 & 2.0 & 16& 3.0 & 2.0 & 28 & 5.5 & 5.5 & 40 & 2.25 & 2.25 & 52 & 2.62 & 2.62 \cr
5 & 1.0 & 1.0 & 17& 8.0 & 8.0 & 29 & 4.5 & 4.5 & 41 & 6.1 & 6.1 & 53 & 3.82 & 3.82 \cr
6 & 2.5 & 2.5 & 18& 8.0 & 2.0 & 30 & 5.75 & 5.75 & 42 & 5.6 & 5.6 & 54 & 4.2 & 4.2 \cr
7 & 3.0 & 3.0 & 19& 2.0 & 8.0 & 31 & 6.25 & 6.25 & 43 & 5.0 & 5.9 & 55 & 4.3 & 4.3 \cr
8 & 0.2 & 0.2 & 20 & 7.0 & 7.0 & 32 & 6.75 & 6.75 & 44 & 6.4 & 6.4 & 56 & 4.42 & 4.42 \cr
9 & 3.5 & 3.5 & 21 & 5.0 & 5.0 & 33 & 5.25 & 5.25 & 45 & 5.1 & 5.1 & 57 & 1.62 & 3.81 \cr
10& 4.0 & 4.0 & 22 & 9.0 & 9.0 & 34 & 4.25 & 4.25 & 46 & 4.6 & 4.6 &    &      &      \cr
11& 0.1 & 0.1 & 23 & 10.0 & 10.0 & 35 & 4.75 & 4.75 & 47 & 4.1 & 4.1 &    &      &    \cr
12& 2.0 & 1.0 & 24 & 7.5 & 7.5 & 36 & 3.75 & 3.75 & 48 & 4.35 & 4.35 &    &      &
\end{tabular}
\caption{\label{tab:para}Variational parameters for the radial functions where $n$ 
resembles the profile numbers from the figures in Sec.~IV.}
\end{table}


\begin{thebibliography}{99}

\bibitem{Copeland:2011dx}
  E.~J.~Copeland, L.~Pogosian, T.~Vachaspati,
  Class.\ Quant.\ Grav.\  {\bf 28} (2011) 204009.

\bibitem{Hindmarsh:2011qj}
  M.~Hindmarsh,
  Prog.\ Theor.\ Phys.\ Suppl.\  {\bf 190} (2011) 197.

\bibitem{Copeland:2009ga}
  E.~J.~Copeland, T.~W.~B.~Kibble,
  Proc.\ Roy.\ Soc.\ Lond.\  A {\bf 466}, 623 (2010).

\bibitem{Nielsen:1973cs}
  H.~B.~Nielsen, P.~Olesen,
  Nucl.\ Phys.\  B {\bf 61} (1973) 45.

\bibitem{Hindmarsh:2016lhy}
  M.~Hindmarsh, K.~Rummukainen, D.~J.~Weir,
  arXiv:1607.00764 [hep-th].

\bibitem{Achucarro:1999it}
  A.~Achucarro, T.~Vachaspati,
  Phys.\ Rept.\  {\bf 327}, 347 (2000).

\bibitem{Kibble:2015twa}
  T.~W.~B.~Kibble, T.~Vachaspati,
  J.\ Phys.\ G {\bf 42} (2015)  094002 (2015).

\bibitem{Naculich:1995cb}
  S.~G.~Naculich,
  Phys.\ Rev.\ Lett.\  {\bf 75} (1995) 998.

\bibitem{Klinkhamer:2003hz}
  F.~R.~Klinkhamer, C.~Rupp,
  J.\ Math.\ Phys.\  {\bf 44} (2003) 3619.

\bibitem{Stojkovic}
  G.~Starkman, D.~Stojkovic, T.~Vachaspati,
  Phys.~Rev.~D~\textbf{65} (2002) 065003.\\
  G.~Starkman, D.~Stojkovic, T.~Vachaspati,
  Phys.~Rev.~D~\textbf{63} (2001) 085011. \\
  D.~Stojkovic,
  Int.~J.~Mod.~Phys.~A~\textbf{16} (2001) 1034.

\bibitem{Groves:1999ks}
  M.~Groves, W.~B.~Perkins,
  Nucl.\ Phys.\  B {\bf 573} (2000) 449.

\bibitem{Bordag:2003at}
  M.~Bordag, I.~Drozdov,
  Phys.\ Rev.\  D {\bf 68} (2003) 065026.

\bibitem{Baacke:2008sq}
  J.~Baacke, N.~Kevlishvili,
  Phys.\ Rev.\  D {\bf 78} (2008) 085008.

\bibitem{Graham:2009zz}
  N.~Graham, M.~Quandt, H.~Weigel,
  Lect.\ Notes Phys.\  {\bf 777} (2009) 1.

\bibitem{Weigel:2015lva}
  H.~Weigel, M.~Quandt, N.~Graham,
  Mod.\ Phys.\ Lett.\ A {\bf 30} (2015) 1530022\,.

\bibitem{Weigel:2010pf}
  H.~Weigel, M.~Quandt,
  Phys.\ Lett.\  B {\bf 690} (2010) 514.

\bibitem{Weigel:2009wi}
  H.~Weigel, M.~Quandt, N.~Graham, O.~Schr\"oder,
  Nucl.\ Phys.\  B {\bf 831} (2010) 306.

\bibitem{Graham:2011fw}
  N.~Graham, M.~Quandt, H.~Weigel,
  Phys.\ Rev.\ D {\bf 84} (2011) 025017.

\bibitem{Farhi:2001kh}
  E.~Farhi, N.~Graham, R.~L.~Jaffe, H.~Weigel,
  Nucl.\ Phys.\ B {\bf 630} (2002) 241\,.

\bibitem{Klinkhamer:1997hw}
  F.~R.~Klinkhamer, C.~Rupp,
  Nucl.\ Phys.\ B {\bf 495} (1997) 172\,.

\bibitem{Graham:2003ib}
  N.~Graham, R.~L.~Jaffe, V.~Khemani, M.~Quandt, O.~Schr\"oder, H.~Weigel,
  Nucl.\ Phys.\ B {\bf 677} (2004) 379.

\bibitem{Farhi:2000ws}
  E.~Farhi, N.~Graham, R.~L.~Jaffe, H.~Weigel,
  Nucl.\ Phys.\ B {\bf 585} (2000) 443.

\bibitem{Farhi:2000gz}
  E.~Farhi, N.~Graham, R.~L.~Jaffe, H.~Weigel,
  Nucl.\ Phys.\ B {\bf 595} (2001) 536.

\bibitem{Pasipoularides:2000gg}
  P.~Pasipoularides,
  Phys.\ Rev.\ D {\bf 64} (2001) 105011,~
  hep-th/0502238.

\bibitem{Schroder:2007xk}
  O.~Schr\"oder, N.~Graham, M.~Quandt, H.~Weigel, 
  J.\ Phys.\ A  {\bf 41} (2008) 164049.

\bibitem{Klinkhamer:1994uy}
  F.~R.~Klinkhamer, P.~Olesen,
  Nucl.\ Phys.\ B {\bf 422} (1994) 227\,.

\bibitem{Graham:2006qt}
  N.~Graham, M.~Quandt, O.~Schr\"oder, H.~Weigel,
  Nucl.\ Phys.\ B {\bf 758} (2006) 112\,.

\bibitem{Faulkner:1977aa}
  J.~S.~Faulkner, J. Phys. C:. 
  Solid State Phys.,{\bf 10} (1977) 4661.

\bibitem{Graham:2001iv}
  N.~Graham, R.~L.~Jaffe, M.~Quandt, H.~Weigel,
  Annals Phys.\  {\bf 293} (2001) 240.

\bibitem{Graham:2001dy}
  N.~Graham, R.~L.~Jaffe, M.~Quandt, H.~Weigel,
  Phys.\ Rev.\ Lett.\  {\bf 87} (2001) 131601.

\bibitem{Newton:1982qc}
R. G. Newton, {\it Scattering Theory of Waves and Particles}
Springer, New York (1982).

\bibitem{Chadan:1977pq}
K. Chadan, P. Sabatier, {\it Inverse Problems in Quantum Scattering Theory}
Springer, New York (1977).

\bibitem{Ripka:1987ne}
  G.~Ripka, S.~Kahana,
  Phys.\ Rev.\  D {\bf 36} (1987) 1233.

\bibitem{Hartmann:1994ai}
  J.~Hartmann, F.~Beck, W.~Bentz,
  Phys.\ Rev.\  C {\bf 50} (1994) 3088.

\bibitem{laug1999}
      E. Anderson {\it el al.}, 
      ``LAPACK Users' Guide'' (1999),
      Soc. for Industrial \& Applied Mathematics,
      ISBN~0-89871-447-8.

\bibitem{Diakonov:1993ru}
  D.~Diakonov, M.~V.~Polyakov, P.~Sieber, J.~Schaldach, K.~Goeke,
  Phys.\ Rev.\ D {\bf 49} (1994) 6864.

\bibitem{Weigel:2010zk}
  H.~Weigel, M.~Quandt, N.~Graham,
  Phys.\ Rev.\ Lett.\  {\bf 106} (2011) 101601.

\bibitem{AS}
  M. Abramowitz, I. Stegun (eds.),
  {\it Handbook of mathematical functions}, Dover, New York (1968).
  
\end{thebibliography}
\end{document}